\documentclass[aps,prb,reprint,longbibliography,superscriptaddress]{revtex4-2} 
\usepackage{mathrsfs}
\usepackage{textcomp, gensymb}
\usepackage{amsmath,gensymb}
\usepackage{amsfonts}
\usepackage{amssymb}
\usepackage{amsthm}
\usepackage{mathtools}
\usepackage{graphicx}
\usepackage{natbib}
\usepackage{xcolor}
\usepackage{hyperref}
\usepackage{bm}
\usepackage[caption=false]{subfig}
\usepackage{verbatim}
\usepackage{braket}
\usepackage{physics}
\usepackage{nicefrac}
\usepackage{ulem}

\newcommand{\Hmax}{H_{\mathrm{max}}}
\newcommand{\jbar}{\bar{\pmb{\j}}}
\newcommand{\Fl}{\bar{\pmb{F}}_{\scriptscriptstyle L}}
\begin{document}
\title{Nonreciprocity of supercurrent along applied magnetic field}

\author{Filippo Gaggioli}
\email[Email: ]{gfilippo@mit.edu}
\affiliation{Department of Physics, Massachusetts Institute of Technology, Cambridge, MA-02139,USA}

\author{Yasen Hou}
\affiliation{Francis Bitter Magnet Laboratory \& Plasma Science and Fusion Center, Massachusetts Institute of Technology, Cambridge, MA-02139, USA}

\author{Jagadeesh S. Moodera}
\affiliation{Francis Bitter Magnet Laboratory \& Plasma Science and Fusion Center, Massachusetts Institute of Technology, Cambridge, MA-02139, USA}
\affiliation{Department of Physics, Massachusetts Institute of Technology, Cambridge, MA-02139,USA}

\author{Akashdeep Kamra}
\affiliation{Department of Physics and Research Center OPTIMAS, Rheinland-Pf\"alzische Technische Universit\"at Kaiserslautern-Landau, 67663 Kaiserslautern, Germany} 
\affiliation{Condensed Matter Physics Center (IFIMAC) and Departamento de Física Teórica de la Materia Condensada, Universidad Autónoma de Madrid, E-28049 Madrid, Spain}

\begin{abstract}
Nonreciprocal currents arise in a broad range of systems, from magnons and phonons to supercurrents, due to an interplay between spatial and temporal symmetry breakings. These find applications in devices, such as circulators and rectifiers, as well as in probing the interactions and states that underlie the nonreciprocity. An established symmetry argument anticipates emergence of nonreciprocal currents along a direction perpendicular to the applied magnetic field that breaks the time-reversal symmetry. Here, motivated by recent experiments, we examine the emergence of nonreciprocity in vortex-limited superconducting critical currents along an applied magnetic field. Employing London's equations for describing the Meissner response of a superconducting film, we find that an additional symmetry breaking due to a preferred vortex axis enables nonreciprocal critical currents along the applied magnetic field, consistent with the so far unexplained experimental observation. Building on our concrete theoretical model for supercurrents, we discuss a possible generalization of the prevailing symmetry consideration to encompass nonreciprocal currents along the time-reversal symmetry breaking direction. 
\end{abstract}
\maketitle

\section{Introduction}

When forward and backward directions can be distinguished in a system, nonreciprocal behavior can manifest via, for example, different resistances or currents along these opposite directions~\cite{Tokura_2018,Nagaosa_2024}. A prototypical example is a p-n junction diode in which a simple identification of p-type and n-type subsystems admits different backward and forward current flows. Thus, in such systems divisible into two ``lumped'' subsystems, the spatial-inversion breaking along a direction $\hat{\pmb{n}}$ alone can be sufficient to realize nonreciprocal behavior along this same axis~\cite{Hu_2007,Strambini_2022,Amundsen_2022,Geng_2023}. In extended systems, one may distinguish between forward and backward transport along the axis $\hat{\pmb{n}} \times \hat{\pmb{h}}$, where $\hat{\pmb{h}}$ is the direction of time-reversal symmetry breaking via an applied magnetic field or similar~\cite{Rikken_2001,Rikken_2002,Edelstein_1990}. Consistent with this principle, nonreciprocal responses of magnons~\cite{Ogawa_2021,Yang_2021,Guckelhorn_2023,Yu_2023}, phonons~\cite{Kuss_2020,Kuss_2022}, and Cooper pairs~\cite{Ando_2020,Baumgartner_2022,Pal_2021,Hou_2023,Nadeem_2023,Cadorim_2024} perpendicular to an applied magnetic field have been observed using a wide range of systems and mechanisms~\cite{Tokura_2018,Nagaosa_2024}. 
In general, these nonreciprocal behaviors have been broadly divided~\cite{Nagaosa_2024} into \emph{intrinsic} diode effects,
originating from the microscopic properties of the material, e.g., a strong spin-orbit coupling resulting in finite momentum Cooper pairing, 
and \emph{extrinsic} diode effects, where the nonreciprocity is a result of the sample geometry.
Intrinsic responses offer a convenient probe for unconventional interactions and states of quantum matter~\cite{Yuan_2022,Daido_2022,He_2022,Ilic_2022},
while extrinsic diodes are very promising for technological applications as they may enable useful devices, such as rectifiers and circulators.
Both these tasks, namely design of devices and effective probing, strongly rely on a general understanding of the symmetries that allow for nonreciprocal responses~\cite{Nagaosa_2024,Zinkl_2022,Moll_2023}.

\begin{figure*}
	\centering
	\includegraphics[width=\textwidth]{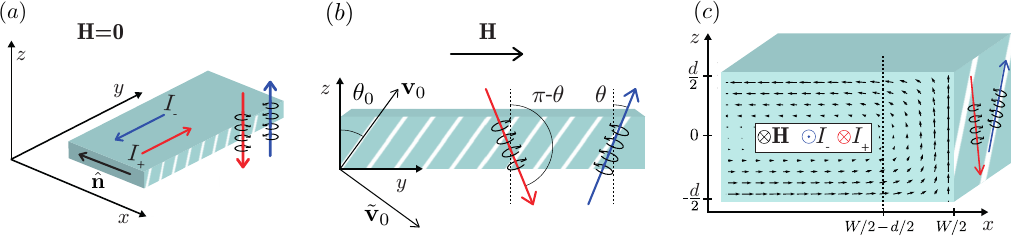}
	\caption{Schematic depiction of the vortex mechanism underlying the SDE in an in-plane magnetic field $\mathbf{H}$ parallel to the applied bias current $\mathbf{I}$. 
	$(a)$ For $\mathbf{H} = 0$, vortices penetrate inside the superconductor with their axis out-of-plane. For a sample with asymmetry along the $\hat{\pmb{n}}$ direction, the right surface barrier, for example, is weaker and the vortices preferably penetrate from this edge with their axis (anti-)parallel to the $z-$axis for negative (positive) currents, thereby determining $I^\pm_c(0)$. 
	$(b)$ At finite $\mathbf{H}$, the vortices tilt in the direction of the magnetic field and form an angle $\theta$ ($\pi - \theta$) with the $z-$axis for negative (positive) currents. When the surface barrier is angular dependent  and strongest (or weakest) along $\mathbf{v}_0$ ($\tilde{\mathbf{v}}_0$), e.g., because of the lithographic process (white stripes), the different barriers experienced by the titled vortices result in unequal critical current densities for positive and negative current flows, giving rise to the SDE.
	$(c)$ The current density across a $xz-$section of the thin film close to the right edge. This Meissner response to $\mathbf{H}$ is given by Eq.\ \eqref{eq:Meissner_edge_z} (for the $z-$component) and more broadly discussed in Appendix \ref{appendix_currents}. 
	}
\label{fig1}
\end{figure*}

It therefore came as a surprise when unequal superconducting critical currents, a phenomenon dubbed the superconducting diode effect (SDE)~\cite{Nadeem_2023,Ando_2020}, were observed in thin film superconductors with negligible spin-orbit coupling subjected to a magnetic field {\it parallel} to the current flow direction~\cite{Hou_2023}, because the observation defied the symmetry-based expectation stated above. 
The phenomenon of SDE has gained a renewed interest~\cite{Edelstein_1990,Vodolazov_2005,Vodolazov_2005B,Cerbu_2013} and focus in the recent years with its observation via a broad range of systems and mechanisms~\cite{Ando_2020,Baumgartner_2022,Pal_2021,Hou_2023,Nadeem_2023,Cadorim_2024}. 
We here focus on thin films of a nominally centrosymmetric superconductor, where the time-reversal and spatial-inversion symmetries are broken, respectively, by an applied magnetic field and by inequivalent vortex surface barriers on the two sides~\cite{Plourde_2001,Vodolazov_2005,Hope_2021,Gaggioli_2024}, possibly due to defects and geometric features introduced during the lithography process~\cite{Cerbu_2013,Hou_2023}. 
The critical current in such thin films is typically determined by the Bean-Livingston vortex surface barrier~\cite{Bean_1964}. The current which is large enough to exert sufficient Lorentz force on the vortices to overcome the weakest surface barrier becomes the critical value~\cite{Shmidt_1970,Shmidt_1970B,Maksimova_1998,Hope_2021} (Fig.~\ref{fig1}). The consequent vortex-mediated SDE has been investigated and understood for applied magnetic fields perpendicular to the current flow direction~\cite{Cerbu_2013,Hou_2023,Suri_2022,Gutfreund_2023,Chahid_2023}, consistent with the prevailing symmetry argument above.

In this work, we examine the critical currents and SDE in a conventional centrosymmetric superconductor film subjected to a magnetic field parallel to the current flow direction, thereby going beyond the conventional symmetry consideration. 
Considering the surface barrier mechanism typical of superconducting thin films \cite{Gaggioli_2024},
we find that the critical current is determined by the penetration of vortices whose axis is (weakly) tilted in the applied field direction.
A surface barrier that is maximal for vortices with a finite tilt angle then gives rise to different critical currents in the forward and backward directions [Fig.~\ref{fig1}$\,(b)$]. Such an angular dependence for the vortex barrier may result from, for example, (controllable) geometrical defects induced by lithographic preparation of the film. Our theoretical results fit the experimental data semi-quantitatively. We further discuss a possible generalization of the prevailing symmetry argument that finds nonreciprocity along the direction $\hat{\pmb{n}} \times V_0 \hat{\pmb{h}}$, where $V_0$ is a tensor characterizing the vortex barrier angular anisotropy. This reduces to the standard $\hat{\pmb{n}} \times \hat{\pmb{h}}$ framework when the vortex barrier has no angular dependence.

\section{Vortex-limited critical current}

In a wide range of superconducting films, the critical supercurrent is determined by the presence of a steep surface barrier preventing the penetration of vortices inside the sample \cite{Shmidt_1970,Shmidt_1970B,Maksimova_1998,Clem_2010,Clem_2011,Gaggioli_2024}. This barrier is first overcome when the average Lorentz force density $\Fl$ acting on the vortices -- the axis of which can be slightly tilted away from the out-of-plane direction, see Appendix \ref{app:vortex_tilting} --
reaches a critical value, associated to an equivalent critical current density $j_s$. 
For a thin film of width $-W/2\leq x\leq W/2$ that extends indefinitely along $y$, this critical condition reads
\begin{equation}\label{eq:lorentz_force}
\Fl\left(\pm\frac{W}{2}\right) =  \left(\frac{\phi_0}{c}\right)\jbar\times\pmb{\hat{e}}_v= \mp\left(\frac{\phi_0}{c}\right)j_s\,\hat{\pmb{x}},
\end{equation}
for the vortex axis $\pmb{\hat{e}}_v=(0,\sin\theta,\cos\theta)$ that maximizes the Lorentz force working against the surface barrier~\cite{Shmidt_1970,Hope_2021}. Here, we denote the current density $\pmb{j}$ averaged over the coherence length $\xi$ by $\jbar$, $\phi_0 = hc/2e$ is the flux quantum. The task at hand is therefore to calculate the average current density $\jbar$ flowing at distances $\lesssim \xi$ from the superconductor edges.
Vortices with axis $\pmb{\hat{e}}_v\perp \jbar$ will then be the first to fulfill the condition \eqref{eq:lorentz_force}, thereby determining $I_c(H)$. Here, we disregard the elastic energy contribution due to the vortex tilting and the effect of the boundary condition for the current at the top and bottom film surfaces. A detailed consideration of these effects is presented in the Appendix~\ref{app:vortex_tilting} and supports the results presented below.

In the absence of trapped vortices, the current density $\pmb{j}$ inside the superconductor is determined self-consistently by the interplay of 
the external field $\pmb{H}$ and the self-field produced by the current. 
In thin films with thickness $d$ and width $W$ much smaller than the London and Pearl lengths $\lambda$ and $\lambda_\perp = 2\lambda^2/d$ \cite{Pearl_1964}, respectively, the self-field is negligible \cite{Gaggioli_2024} and the magnetic field is approximately constant inside and outside the superconductor. 
In this case, the current density distribution is found from the London equation~\cite{Shmidt_1970,Gaggioli_2024}
\begin{equation}\label{eq:london_curl}
   \pmb{\nabla}\times\pmb{j}=-\frac{c}{4\pi\lambda^2}\pmb{H},
\end{equation}
with the additional condition that the integral $\int \!j_y \,\mathrm{d}x\,\mathrm{d}z$ is equal to the applied bias current $\pmb{I}\parallel\hat{\pmb{y}}$.

The solution to Eq.\ \eqref{eq:london_curl} depends on the magnetic field, the bias current and, via the vanishing of the current density $j_\perp$ at the boundaries, on the precise film geometry.
For a perpendicular magnetic field along the z axis, 
Eq.\ \eqref{eq:london_curl} yields $x$-dependent current density along $y$,
\begin{equation}\label{eq:Meissner_perp}
j_y(x) =\frac{I}{dW} - \frac{c}{4\pi\lambda^2} Hx.
\end{equation}
Following the principle described via Eq.\ \eqref{eq:lorentz_force}, vortices enter the superconductor when the condition $j_y(x) = j_s$ is first met on either side of the film ($x = \pm W/2$): at zero magnetic field (external bias), this happens when the external bias (magnetic field) reaches \cite{Gaggioli_2024}
\begin{equation}\label{eq:perp_field}
I_0 \equiv j_s\, dW,\quad H_s \equiv \frac{8\pi\lambda^2}{cW} j_s\sim \frac{\phi_0}{\xi W}.
\end{equation}
%

In the parallel-to-current field scenario presented in Fig.\ \ref{fig1}$\,(b)$, on the other hand, the bias current determines $j_y \!=\!\bar{\j}_y \!=\! I/d\,W$,
while the screening currents $\propto H$ flow in the $xz$-plane [Fig.\ \ref{fig1}$\,(c)$] with boundary conditions $j_x(x\!=\!\pm W/2, z) \!=\! j_z(x, z\!=\!\pm d/2) = 0$.
The evaluation of these Meissner currents is detailed in Appendix \ref{appendix_currents}.
For a small aspect ratio $d/W \ll 1$, we then find that
\begin{align}\label{eq:Meissner_edge_z}
j_z (x,z) \approx \pm\frac{c}{4\pi\lambda^2}Hd\left(\sum_{n=0}^{\infty} \frac{(-1)^{n}}{(k_nd/2)^2}e^{-k_n \Delta x}\cos{k_n z}\right),
\end{align}
with $k_n = (n + 1/2)\,2\pi/d$ and $\Delta x = |W/2 \mp x|$ the distance from the edges.
As expected, $j_z$ grows rapidly at distances $\lesssim d/2$ from the edges, where the effect of the boundary conditions $\pmb{j}\perp\hat{\pmb{x}}$ is important.

To average $j_z$ over the size $\xi$ of the vortex core, we take advantage of the rapid decay of the sum in Eq.\ \eqref{eq:Meissner_edge_z} and consider only the contribution of the $n=0$ term to $\bar{\j}_z$. This approximation is valid for thicknesses $d\sim \xi$ (and even somewhat larger: e.g., the ratio between the $n=1$ and $n=0$ term is of the order of one percent for $d = 3\xi$).
In this limit, the average current density is given by
\begin{align}\label{eq:jbar_z}
\bar{\j}_z \approx \pm\frac{2d}{\xi}\left(\frac{2}{\pi^2}\right)^2\!\frac{c}{4\pi\lambda^2}Hd.
\end{align}
As expected, Eq.\ \eqref{eq:jbar_z} vanishes in the limit $d\to 0$, as the Meissner response of the superconductor to in-plane fields becomes negligible in the two-dimensional limit.
 
Using that $\hat{\pmb{e}}_v\!\perp\!\jbar$ for maximal Lorentz force and inserting the expressions for $\bar{\j}_y,\,\bar{\j}_z$, the condition \eqref{eq:lorentz_force} then provides an equation for $I_c(H)$,
\begin{equation}\label{eq:crit_condition}
\sqrt{\left(\frac{I_c(H)}{dW}\right)^2 + \left(\frac{2d}{\xi}\left(\frac{2}{\pi^2}\right)^2\!\frac{c}{4\pi\lambda^2}Hd\right)^2} = j_s.
\end{equation}
This finally yields the field dependence of the critical current
\begin{align}\label{eq:I_c_H}
I_c(H) = I_0\sqrt{1 - \left(\frac{H}{H^\parallel_s}\right)^2},
\end{align}
with the characteristic field scale
\begin{align}\label{eq:H_s_parallel}
H_s^\parallel= \left(\frac{\pi}{2}\right)^4\left(\frac{\xi W}{d^2}\right)H_s\sim \left(\frac{\pi}{2}\right)^4\frac{\phi_0}{d^2},
\end{align}
that is directly related to the sample penetration field $H_s$ \eqref{eq:perp_field} in a perpendicular-to-film geometry \cite{Gaggioli_2024}.
Equations \eqref{eq:I_c_H} and \eqref{eq:H_s_parallel} show two interesting features that are in marked constrast with the $I_c(H)$ of the superconducting thin film in a perpendicular magnetic field.
First, the field dependence is quadratic and not linear as for the perpendicular case \cite{Maksimova_1998,Gaggioli_2024,Plourde_2001}. 
Second, the magnetic scale $H_s^\parallel$ is much larger than the corresponding $H_s$, as the prefactor in Eq.\ \eqref{eq:H_s_parallel} is of the order of $\sim 10^{4}$ for $\xi/d\sim 1$ and a typical aspect ratio $d/W\sim 10^{-3}$.
It then follows that an experiment measuring changes in the critical current for perpendicular fields in the order of Gauss should observe similar variations in $I_c(H)$ for parallel fields in the range of Tesla. Notice that the field scale $H_s^\parallel$ diverges as $d\to 0$, as a consequence of the vanishing of the in-plane Meissner response \eqref{eq:jbar_z}. This indicates the natural limit of applicability of our theory, which relies on the finite thickness of the superconducting film and therefore breaks down in the two-dimensional limit $d\to0$.

Having found the critical current \eqref{eq:I_c_H}, we fix $\bar{\j}_x= I_c(H)$ and use that $\hat{\pmb{e}}_v\!\perp\!\jbar$ to evaluate the vortex tilt angles $\theta^{\scriptscriptstyle(L,R)}$ on the left and right edge
\begin{equation}\label{eq:opt_angles}
\left(\sin\theta^{\scriptscriptstyle(L,R)},\cos\theta^{\scriptscriptstyle(L,R)}\right) = \left(\frac{H}{H^\parallel_s}, \pm\,\frac{I}{|I|}\sqrt{1 - \left(\frac{H}{H^\parallel_s}\right)^2}\right).
\end{equation}
For finite fields $H\ll H_s^\parallel$, the vortices are slightly tilted away from the out-of-plane axis. Moreover, 
 as shown in Fig.\ \ref{fig1}$(b)$ and in agreement with the pseudovector properties of the magnetic field, the tilt angles at the two edges are exchanged upon switching the direction of the bias current $I$, such that $\theta^{\scriptscriptstyle R} \to \theta^{\scriptscriptstyle L}=\pi - \theta^{\scriptscriptstyle R}$.

\section{Superconducting diode effect}\label{sec:diode}
%
%

Let us now discuss the nonreciprocal transport properties of the parallel-to-current field setup discussed above~\cite{Hou_2023}. 
We first recapitulate the SDE in a field perpendicular to the film plane, as this provides a useful comparison.
The SDE, in the perpendicular case, is realized when the critical current densities $j_s^{\scriptscriptstyle (L,R)}$ on the left and right edge of the superconductor are not identical,
\begin{equation}\label{eq:simple_js_edges}
j_{s}^{\scriptscriptstyle(L,R)} = j_{s,0} \mp \Delta j_{s} \left(\hat{\pmb{n}}\cdot\hat{\pmb{x}}\right),
\end{equation}
with $\Delta j_{s}$ assumed positive, without a loss of generality, and the asymmetry vector $\hat{\pmb{n}}=\left(-1,0,0\right)$ shown in Fig.\ \ref{fig1}$\,(a)$.
As a result, $I_c(H)$ reaches its maximum at the peak field \cite{Gaggioli_2024}
\begin{equation}\label{eq:Hmax_perp}
\frac{\Hmax}{H_{s,0}}  = \frac{\Delta j_{s}}{j_{s,0}}\left[(\hat{\pmb{n}}\times \hat{\pmb{h}})\cdot\hat{\pmb{I}}\right],
\end{equation}
which in turn determines the magnitude of the SDE.

In a parallel in-plane field, the magnitude of the current density $\jbar$ at criticality is constant and determined by the weakest surface barrier.
Vortices then always enter from the same edge, suggesting that no SDE can be realized
in agreement with the fact that $\hat{\pmb{n}} \times \hat{\pmb{h}}=0$.
To understand the parallel-field SDE, we consider an additional kind of symmetry breaking, this time at the level of the individual surface barriers, that allows the system to distinguish between opposite signs of the bias current. We take into account a dependence of the surface barrier on the vortex tilt angle $\theta$
\begin{equation}\label{eq:simple_js_theta}
j_{s}^{\scriptscriptstyle(L,R)}(\theta) = j_s^{\scriptscriptstyle(L,R)} + \delta j_{s}\cos2\left(\theta_0 -\theta\right),
\end{equation}
that may result, for example, from columnar tracks left by the litographic process, represented as white stripes in Fig.\ \ref{fig1}. In assuming the form of this angular dependence, time-reversal symmetry requires the surface barrier to be the same for vortices with opposite fluxes. 
Considering reflections across the $xz-$plane, we then expect that the vortices will experience a different surface barrier \eqref{eq:simple_js_theta} as $\theta\to\pi - \theta$ and $I\to -I$, giving rise to the SDE as long as $\theta_0\neq 0, \pi/2$.
In agreement with the time-reversal symmetry, SDE vanishes for $H=0$ as the $\cos2\theta$ dependence ensures that $j_s(\theta)$ remains the same when $\theta$ is changed from $0$ to $\pi$.

 In evaluating the critical condition \eqref{eq:lorentz_force} with $j_s(\theta)$ given by Eq.~\eqref{eq:simple_js_theta}, we assume that the angular dependence is weak, i.e., $\delta j_{s} \ll j^{\scriptscriptstyle(L,R)}_{s}$, and
much weaker than the difference between $j_{s}^{\scriptscriptstyle L}$ and $j_{s}^{\scriptscriptstyle R}$, i.e., $\delta j_{s} \ll \Delta j_{s}$. 
This corresponds to the limit where vortex penetration happens from the weak (assumed right here) edge only, such that $j_{s}^{\scriptscriptstyle R}(\theta)$ alone determines the critical current $I_c(H)$ while $j_{s}^{\scriptscriptstyle L}(\theta)$ does not play any role.
In what follows, we will therefore neglect the $(L, R)$ indices unless necessary.

With the right tilt angle $\theta$ given by Eq.\ \eqref{eq:opt_angles}, the field dependence of the surface barrier \eqref{eq:simple_js_theta} can be immediately found
(we neglect terms $\propto(\delta j_s/j_{s})^2$),
\begin{align}\label{eq:j_s(H)}
j_{s}(H) &= j_{s}\left[1 + \frac{\delta j_{s}}{j_{s}}\cos\left(2\theta_0 \pm 2H/H_s^\parallel\right)\right],
\end{align}
where the plus and minus signs refer to positive and negative bias currents.
To determine $I_c(H)$ from Eq.\ \eqref{eq:lorentz_force}, we now plug in \eqref{eq:j_s(H)} and expand the cosine term around $2\theta_0$ while neglecting terms $\propto(\delta j_s/j_{s})^2$ and $\propto(\delta j_s/j_{s})(H/H^\parallel_{s})^2$ leading to
\begin{equation}\label{eq:I_c_H_shifted}
I^\pm_c(H) \approx I_{\max}\sqrt{1
  - \left(\frac{H - H^\pm_{\max}}{H^\parallel_{s}}\right)^2},
\end{equation}
where we introduced the peak field
\begin{equation}\label{eq:H_max_parallel}
\Hmax^\pm = \mp \left(2\sin2\theta_0\frac{\delta j_{s}}{j_{s}}\right)H^\parallel_{s},
\end{equation}
and the peak current
\begin{align}\label{eq:H_s_parallel_eff}
I_{\max} = I_0\sqrt{1 + \cot(2\theta_0)\frac{|H^\pm_{\max}|}{H_{s}^\parallel}}.
\end{align}

\begin{figure}
        \centering
        \includegraphics[width = 1.\columnwidth]{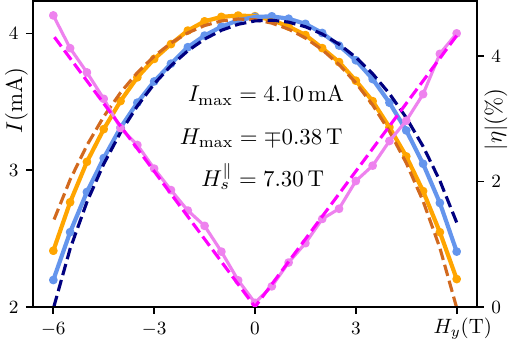}
	\caption[Critical current]{Experimental data from Ref.\ \onlinecite{Hou_2023} (circles and solid lines) for the critical current $I^\pm_c(H)$ (orange and blue) and the SDE efficiency $|\eta(H)|$ (pink) for a superconducting vanadium thin film in a parallel in-plane magnetic field.
	Fitting Eqs.\ \eqref{eq:I_c_H_shifted} and \eqref{eq:eta_H} from our theoretical model (dashed lines), we find very good agreement for the indicated choice of parameters. The field $H_s^\parallel\approx 7.30\,\text{T}$ obtained from fitting agrees well with the prediction $H_s^\parallel\sim 9\,\text{T}$ from Eq.\ \eqref{eq:H_s_parallel}, using the experimental values for $H_s, \xi, d$ and $W$ from Ref.\ \onlinecite{Hou_2023}.
	}
    \label{fig:fit}
\end{figure}

The shifted field dependence \eqref{eq:I_c_H_shifted} results from the two-fold contributions of the magnetic field to vortex penetration:
on the one hand, to enhance the Meissner current \eqref{eq:jbar_z} and, on the other, to reinforce the surface barrier \eqref{eq:j_s(H)}.
These effects compensate at the peak field $\Hmax$, which does not depend on the left-right asymmetry $\Delta j_s/j_{s,0}$, as in Eq.\ \eqref{eq:Hmax_perp}, but is fixed entirely by the parameters $\delta j_{s}/j_{s}$ and $\theta_0$ in the limit where $I_c(H)$ is determined by the right edge alone.

As in the case of a perpendicular field,
the peak field $\Hmax$ determines the efficiency $\eta$ of the superconducting diode device
$\eta(H) \equiv (I_c^+(H) - I_c^-(H))/(I_c^+(H) +I_c^-(H))$.
Using Eq.\ \eqref{eq:I_c_H_shifted} and neglecting terms $\propto (H_{\max}/H^\parallel_s)^2$ and $\propto (H_{\max}/H^\parallel_s)(H/H_s^\parallel)^3$, this is found to read
\begin{equation}\label{eq:eta_H}
    \eta(H) \approx \left(\frac{H^+_{\max}}{H_s^\parallel}\right)\frac{H}{H_s^\parallel}.
\end{equation}

Equations \eqref{eq:I_c_H_shifted} and \eqref{eq:eta_H} lend themselves conveniently to a comparison with the experimental results on the parallel-to-current field SDE reported in Ref.\ \onlinecite{Hou_2023}.
Fitting $I_c^\pm(H)$ and $\eta(H)$ to the experimental data, 
we obtain the dashed curves in Fig.\ \ref{fig:fit}. These reproduce the critical currents and the efficiency curve very well for the choice of $I_{\max},\, H_{\max},\, H_s^\parallel$ reported in the plot. The small differences can result from the various approximations employed in our analytic simplifications.
This combination of parameters is consistent with the assumptions underlying Eqs.\ \eqref{eq:I_c_H_shifted} and \eqref{eq:eta_H}, as $(\delta j_{s}/j_{s})\sim(\Hmax^\pm/H^\parallel_{s})\ll 1$.

Finally, we compare the field value $H_s^\parallel = 7.30\,\text{T}$ estimated from the fit in Fig.\ \ref{fig:fit} to our theoretical prediction \eqref{eq:H_s_parallel}. 
Using the experimental value $H_s \approx 11\, \text{Oe}$ for the perpendicular penetration field and taking $\xi \approx 11\,\text{nm},\, d \approx 8\,\text{nm}$ and $W = 8\,\mu\text{m}$ from Ref.\ \onlinecite{Hou_2023}, we find
that $H_s^\parallel \sim 9\,\text{T}$, in satisfactory agreement with the result of our fit.

\section{Generalized symmetry indicator for nonreciprocity}

Building upon the insights gained from our theoretical model of the SDE in a magnetic field applied along the current flow direction~\cite{Hou_2023}, we now discuss an appropriate generalization of $\hat{\pmb{n}} \times \hat{\pmb{h}}$ as the direction of nonreciprocity. The crucial addition comes from the angular dependence of the vortex surface barrier. Our model suggests that we should find SDE whenever the vortex tilting due to the applied magnetic field causes the vortices to experience different surface barriers for forward and backward current flows. This general idea supersedes the simple case of applied magnetic field along the current direction, as we now formulate.

As discussed below Eq.\ \eqref{eq:simple_js_theta}, the surface barrier needs to satisfy time-reversal symmetry and thus remain the same when $\theta \to \theta + \pi$. 
Differently from the case of the left-right asymmetry \eqref{eq:simple_js_edges}, the symmetry-breaking associated with the angular dependence $j_s(\theta)$ cannot be parametrized by a single vector because the scalar product with the vortex axis would not be invariant when $\hat{\pmb{e}}_v \to  -\hat{\pmb{e}}_v$ under the time-reversal operation.
On the other hand, the quadratic form determined by the tensor quantity $V_0 = \pmb{v}_0\pmb{v}_0^T  - \tilde{\pmb{v}}_0\tilde{\pmb{v}}_0^T$, built from the 
vectors $\pmb{v}_0 = (0, \sin\theta_0,\cos\theta_0)$ and $\tilde{\pmb{v}}_0 = (0, \cos\theta_0, -\sin\theta_0)$ that constitute the symmetry axes of the angular dependence (see Eq.~\eqref{eq:simple_js_theta} and Fig.\ \ref{fig1}$\,(b)$), respects time-reversal symmetry.
In terms of $V_0$, the surface barrier \eqref{eq:simple_js_theta} reads
\begin{equation}\label{eq:simple_js_theta_tensor}
j_{s}^{\scriptscriptstyle(L,R)}(\theta) = j_{s}^{\scriptscriptstyle(L,R)} + \delta j_{s}\left(\hat{\pmb{e}}_v\,
 V_0\,\hat{\pmb{e}}_v\right),
\end{equation}
such that Eqs.\ \eqref{eq:H_max_parallel} and \eqref{eq:eta_H} take the simple form
\begin{equation}\label{eq:Hmax_par_tensor}
\eta(H)\propto\frac{\Hmax}{H_s^\parallel}  = \frac{2\delta j_s}{j_s}\left[(\hat{\pmb{n}}\times V_0\,\hat{\pmb{h}})\cdot\hat{\pmb{I}}\right].
\end{equation}
%

The peak field \eqref{eq:Hmax_par_tensor} is maximal when $\hat{\pmb{h}}$ is parallel to $\hat{\pmb{I}}$, and vanishes when they are orthogonal. 
The structure of Eqs.\ \eqref{eq:simple_js_theta_tensor} and \eqref{eq:Hmax_par_tensor}, however, suggests that this is just a special case. Considering now edges that are tilted by an angle $\phi$ away from the $z$-axis in the $xz$-plane, such that $\left(\pmb{v}_0,\tilde{\pmb{v}}_0\right)\to R_y(\phi)\left(\pmb{v}_0,\tilde{\pmb{v}}_0\right)$ with the matrix $R_y(\phi)$ describing rotations about the $y$-axis, we find that the tensor $V_0$ transforms as $R_y(\phi)V_0 R^T_y(\phi)$.
Inserting this into Eq.\ \eqref{eq:Hmax_par_tensor} then allows to evaluate $\Hmax$ for an arbitrary direction of the in-plane field $\hat{\pmb{h}}=(h_x,h_y, 0)$
\begin{equation}\label{eq:Hmax_inplane_tensor}
\frac{\Hmax}{H_s^\parallel}  = \frac{2\delta j_s}{j_s}\left[h_y\sin2\theta_0 + h_x\sin\phi\cos2\theta_0\right]\cos\phi,
\end{equation}
with the SDE efficiency $\eta(H)$ again $\propto\Hmax/H^\parallel_s$. This further admits finite SDE when magnetic field is applied along an arbitrary in-plane direction, for example orthogonal to the current, and is consistent with the experiments~\cite{Hou_2023} similar to those discussed in Fig.~\ref{fig:fit} when the field is applied along the $x$-axis.

\section{Conclusion}

 Our theoretical treatment of vortex-limited superconducting critical current offers a possible explanation for, and a good agreement with, the experimentally observed SDE in superconducting thin films subjected to in-plane magnetic fields, especially along the current flow direction. 
It further suggests new avenues for a lithographic control of the SDE via the angle $\theta_0$.
Our consequent generalization of the symmetry arguments for observing nonreciprocity offers an important supplement to the existing understanding and may guide similar search for nonreciprocal responses in other non-superconductor systems.

\begin{acknowledgments}

 F.G. is grateful for the financial support from the Swiss National Science Foundation (Postdoc.Mobility Grant No. 222230) and the support of the EU Cost Action CA16218 (NANOCOHYBRI). A.K. acknowledges financial support from the Spanish Ministry for Science and Innovation -- AEI Grant CEX2018-000805-M (through the ``Maria de Maeztu'' Programme for Units of Excellence in R\&D) and grant RYC2021-031063-I funded by MCIN/AEI/10.13039/501100011033 and ``European Union Next Generation EU/PRTR''.
The work at J.S.M Lab was supported by Air Force Office of Sponsored Research (FA9550-23-1-0004 DEF), Office of Naval Research (N00014-20-1-2306), National Science Foundation (NSF-DMR 1700137 and 2218550); the Army Research Office (W911NF-20-2-0061 and DURIP W911NF-20-1-0074) and the Center for Integrated Quantum Materials (NSF-DMR 1231319).

\end{acknowledgments}

\appendix

\section{Meissner currents distribution}\label{appendix_currents}

In this Appendix, we determine the distribution of the screening currents $j_x, j_z$ induced by an in-plane magnetic field applied parallel to the direction of the bias current.
To do so, we treat the $xz-$section of the thin film as two-dimensional superconductor of length $W$ and width $d$, threaded by a perpendicular magnetic field in the $y-$direction, see Fig.\ 1$\,(c)$ in the main text.

We then write the London equation for the current density $\pmb{j}$ in terms of the vector potential $\pmb{A}$ and the order parameter phase $\varphi$,
\begin{equation}\label{eq:london_full}
\pmb{j}=-\frac{c}{4\pi\lambda^2}\left[\pmb{A}-\frac{\phi_0}{2\pi}\pmb{\nabla}\varphi\right]
\end{equation}
with boundary conditions $j_x(x\!=\!\pm W/2, z) \!=\! j_z(x, z\!=\!\pm d/2) = 0$. Taking the curl of Eq.\ \eqref{eq:london_full} reproduces Eq.\ $(2)$ in the main text.

Having chosen the gauge $\pmb{A} = -Hz\,\hat{\pmb{x}}$, we use the incompressibility condition $\pmb{\nabla}\cdot\pmb{j} = 0$ to obtain the Laplace equation $\nabla^2 \varphi = 0$ with boundary conditions $\partial_z\,\varphi(x,z=\pm d/2) = 0$ and $\partial_x\,\varphi(x=\pm W/2,z)=-(2\pi/\phi_0)Hz$.
The solution is then obtained by the method of separation of variables \cite{Jackson_book} and reads (cf. Ref.\ \onlinecite{Clem_2010})
\begin{equation}\label{eq:phase_
OP}
\varphi(x,z) = \frac{8\pi H}{\phi_0 d}\sum_{n=0}^{\infty}\frac{(-1)^n\cosh\left(k_nx\right)\sin\left(k_nz\right)}{k_n^3\cosh\left(k_nW/2\right)}
\end{equation}
with $k_n = (2n + 1)\,\pi/d$.
Eq.\ \eqref{eq:phase_
OP} satisfies the boundary conditions above, as can be proven with the help of Poisson's summation formula.

Taking the gradient of the phase \eqref{eq:phase_
OP} and inserting into Eq.\ \eqref{eq:london_full}, we obtain the screening current distributions shown in Fig.\ 1$(c)$ in the main text. 
For a film with small aspect ratio $d/W \ll 1$, we expand the hyperbolic cosine and find
\begin{align}\label{eq:Meissner_edge_x_app}
j_x &\approx -\frac{c}{4\pi\lambda^2}Hd\left(\frac{z}{d} - \sum_{n=0}^{\infty} \frac{(-1)^n}{(k_nd/2)^2}e^{-k_n \Delta x}\sin{k_n z}\right),\\
j_z &\approx \pm\frac{c}{4\pi\lambda^2}Hd\left(\sum_{n=0}^{\infty} \frac{(-1)^{n}}{(k_nd/2)^2}e^{-k_n \Delta x}\cos{k_n z}\right),\label{eq:Meissner_edge_z_app}
\end{align}
where the plus and minus signs in $j_z$ refer to the right and left halves of the superconductor and $\Delta x = |W/2 \mp x|$ is the distance from the edges.

\section{Vortex tilting}\label{app:vortex_tilting}

\begin{figure}
        \centering
        \includegraphics[width = 1.\columnwidth]{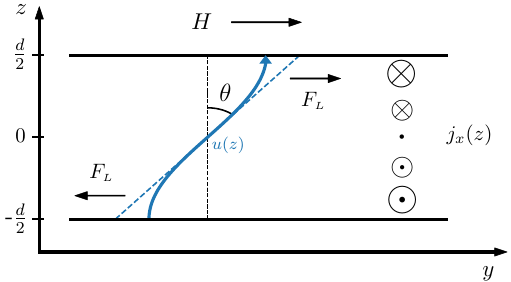}
	\caption[Critical current]{
	The tilted vortex configuration (solid blue line) is determined by the interplay of the Lorentz force density $F_{\scriptscriptstyle L}$ produced by the Meissner current distribution $j_x(z)$, the elastic force resulting from the vortex line tension and the boundary condition that the vortex line is orthogonal to the top and bottom surfaces at $z=\pm d/2$. Because $j_x$ changes sign across $z=0$, the Lorentz force exerts a torque on the vortex line which gives it a finite slope $u'(0)\approx \theta$ away from the top and bottom surfaces, see Eq.\ \eqref{eq:tilt_angle_app}.
	}
    \label{fig:appendix}
\end{figure}

In the main text, we relied on the notion of a tilted vortex line to derive the in-plane-field dependence of the critical current [Eq.~\eqref{eq:I_c_H}] and explain the SDE that is the main focus of this work. For simplicity, we treated vortices as straight lines intersecting the top and bottom surfaces of the superconducting film at the finite angle given by Eq.\ \eqref{eq:opt_angles}. In doing so, we neglected the elastic properties of the vortex line and ignored the role of the boundary condition at the top and bottom surfaces of the superconducting film, which require the vortex line to be orthogonal to these surfaces.
Here, taking this boundary condition into account, we examine the tilted vortex lines in a film with $d \sim \xi$ in order to complement and further support the analysis presented in the main text.

The equilibrium configuration of an isolated vortex is determined by the balance between the elastic forces (resulting from the vortex line tension) and the Lorentz forces (produced by the Meissner currents inside the superconductor) and, in addition, the boundary condition $j_\perp=0$ imposing that the vortex line is orthogonal to sample's surfaces \cite{Brandt_1993_tilt}.
As shown in Fig.\ \ref{fig:appendix}, in the presence of an in-plane field $H\parallel\hat{\pmb{y}}$ this competition yields that the vortex line deviates from the straight out-of-plane configuration and develops a finite tilt angle away from the top and bottom surfaces of the superconductor, while remaining orthogonal to the film plane close to $z = \pm d/2$.

To quantify the deviation of the vortex line from the straight out-of-plane configuration, we introduce the displacement $u(z)$ in the direction of the in-plane magnetic field (see Fig.\ \ref{fig:appendix}) and write the change in the energy of an isolated vortex as \cite{Brandt_1993_tilt,Blatter_1994}
\begin{equation}\label{eq:total_energy_free_vortex}
\Delta E = \int\mathrm{d}z\,\left[\frac{\bar{C}}{2}\left|\frac{\mathrm{d}u(z)}{\mathrm{d}z}\right|^2 - F_{\scriptscriptstyle L}(z)\,u(z)\right],
\end{equation}
with $\bar{C}$ being the elastic constant of the vortex line and $F_{\scriptscriptstyle L}(z) = (\phi_0/c)j_x(z)$ the Lorentz force density at small tilt angles.
Because the Meissner current $j_x(z)$ changes sign across the $z=0$ plane, the Lorentz force exerts a torque on the vortex line that is counteracted by the restoring elastic forces.

To proceed, we re-express the energy \eqref{eq:total_energy_free_vortex} in momentum space. The boundary condition that the vortex line is $\parallel \pmb{\hat{z}}$ at the top and bottom surfaces imposes that $u'(\pm d/2) = 0$, suggesting the Fourier decomposition $u(z) = \sum_n u_n \sin(k_n z)$. Similarly, the odd parity of the Meissner current (cf Eq.\ \eqref{eq:Meissner_edge_x_app}), imposes that $j(z) = \sum_n j_n \sin(k_n z)$, where in both cases $k_n = (2n + 1)\pi/d$ with $n\geq 0$. 
As a result, we obtain that
\begin{equation}
\Delta E = \sum_n \left[\left(\frac{\bar{C}(k_n)}{2}\right)k_n^2 u_n^2 - \left(\frac{\phi_0}{c}\right)j_n u_n\right],
\end{equation}
which allows us to minimize the energy individually for each mode $n$, yielding that 
$u_n = \phi_0\,j_n/c\,\bar{C}(k_n)k_n^2$.

The Fourier transform of Eq.\ \eqref{eq:Meissner_edge_x_app} yields that $j_0 \approx (8/\pi^2)(c H /4\pi\lambda^2)d$ (we ignore for now the contribution of the sum in \eqref{eq:Meissner_edge_x_app}), while the higher modes $j_n$ are suppressed by a factor $\propto (2n + 1)^{-2}$.
In the following, we therefore focus on the dominant $k_0=\pi/d$ mode. 
From Refs.\ \onlinecite{Brandt_1991,Blatter_1994}, we know that the tilt  modulus of an isolated vortex line at short wave-lengths $k^{-1}\sim \xi \ll \lambda$ is given by 
$\bar{C}(k) \approx (\varepsilon_l/\ln(\lambda/\xi)) \ln(1+ 1/k^2\xi^2)^{1/2}\sim (\phi_0/4\pi\lambda)^2$, with the energy line-density of an isolated out-of-plane vortex being $\varepsilon_l = (\phi_0/4\pi\lambda)^2\ln(\lambda/\xi)$.
Plugging this into the expression for $u_0$, we then obtain that $u_0 \sim (8/\pi^4)(Hd^3/\phi_0)$ and, going back to real space, we find that the maximal slope of the vortex is at $z=0$ and reads
 (cf Eq.\ \eqref{eq:H_s_parallel} for $H_s^\parallel$) %
\begin{equation}\label{eq:tilt_angle_app}
u'(0) \approx k_0 u_0= \left(\frac{2}{\pi}\right)^3\frac{H}{\phi_0/d^2}\sim \frac{H}{H_s^\parallel}.
\end{equation}

We have found that, away from the top and bottom surfaces of the superconductor, the vortex line minimizes its energy by tilting in the direction of the in-plane magnetic field.
Because the tilt angle $\theta \approx \tan\theta = u'(0)$ derived in Eq.\ \eqref{eq:tilt_angle_app} in this appendix is of the same order of that in Eq.\ \eqref{eq:opt_angles} in the main text, we conclude that the straight vortex line  approximation that we have used in our work is valid away from $z = \pm d/2$ and yields reasonable results. 
Importantly, the out-of-plane Meissner current $j_z$ in Eq.\ \eqref{eq:Meissner_edge_z} is significant precisely in this region away from the top and bottom surfaces, where they contribute to pushing the vortex against the surface barrier.

We finally comment on the role of the lateral edges after neglecting it in the above derivation.
Close to $x = \pm W/2$ the Meissner current density \eqref{eq:Meissner_edge_x_app} is reduced from the value $j_x = (cH/4\pi\lambda^2)z$ by a factor $\propto \Delta x/d$,
resulting in a corresponding decrease of the Lorentz force density $F_{\scriptscriptstyle L}$ in Eq.\ \eqref{eq:total_energy_free_vortex}.
At distances $\Delta x\sim \xi$ from the lateral edges, however, the presence of an attractive surface barrier equal to $-(\phi_0/4\pi\lambda)^2\ln(\lambda/2\Delta x)$ \cite{Shmidt_1970B}
reduces the energy line density $\varepsilon_l$ of the vortex by a factor $\propto \Delta x/\xi$, resulting in a similar decrease of the tilt modulus $\bar{C}$. 
As a result, the weakening of the Lorentz force density and the softening of the vortex line tend to compensate each other close $\pm W/2$, indicating that our analysis of the tilted vortex configuration remains qualitatively valid also at the edges of the thin film where vortex penetration actually occurs.

\bibliography{refs_film}

\end{document}